\documentstyle[preprint,aps]{revtex}
\draft
\preprint{atom-ph/9510002}
\begin{document}
\title{Quantum uncertainties in coupled harmonic oscillator}
\author{Abir Bandyopadhyay \footnote{Electronic address :
abir@iitk.ernet.in} and Jagdish Rai\footnote{Electronic address :
jrai@iitk.ernet.in}}
\address{Department of Physics, Indian Institute of Technology, \\
 Kanpur 208 016, INDIA}
\date{\today}
\maketitle
\begin{abstract}
    In this paper we analyze the quantum uncertainties and the photon
statistics in the interaction between the two modes of radiation by 
treating them as coupled harmonic oscillator with the motivation of 
controlling quantum properties of one light beam by another. Under the 
rotating wave approximation (RWA) we show that if initially one of the 
modes is coherent and the other one squeezed, then the squeezing and 
non-Poissonianness of the photon statistics can transfer from one mode
to the other. We give a parametric study of these properties depending
upon interaction time and the degree of initial squeezing in one of the
modes.
\end{abstract}

        Squeezed states of quantum systems have been an active area of
interest for more than a decade \cite{walls1}. Nonclassical states of
radiation fields showing squeezing properties have been experimentally
produced in four wave mixing procedures or by passing a coherent beam
through optically nonlinear medium \cite{walls2}. The attempt to generate
squeezed radiation is still on as a challenging technological problem 
\cite{fox}. These quantum features have many potential applications in 
interferometry \cite{hillery} and in noise free transmission of 
information \cite{schu}. \\

    An important goal for the future is the development of all-optical
control and communication systems in which one light beam is controlled
by another lightbeam. These two mode type of interactions can be treated
by two coupled harmonic oscillators interacting through coupling 
\cite{louisell}. As we are interested in quantum properties of light, it
becomes interesting to explore the nature of uncertainties involved in the
interaction of two coupled oscillators. The classical solution of the
coupled harmonic oscillator shows the transfer of energy between the two
oscillators. From this fact, the obvious question arises that what happens
to the other non-classical properties (squeezing, non-Poissonian
statistics). In this paper we calculate the transfer of the non-classical
properties from one mode to the other through the interaction (coupling)
between two radiation fields. For experimental purpose, we consider a dual
channel coupler (fig.1) \cite{love}. It cosists of a pair of optical
waveguides which run in sufficiently close proximity, for a certain
distance, so that coherent coupling takes place between them. The coupling
characteristics is sensitively dependent upon the refractive index
difference between the guides, and in non-centrosymmetric nonlinear
material, such as LiNbO$_3$, this can be controlled through the Pockel's
effect. The coupling characteristics can be controlled by an external
electrical signal.\\

    The Hamiltonian of the coupled harmonic oscillator is given by
(in units of $\hbar$)
\begin{mathletters}
\begin{equation}
H = {H_0} + V
\end{equation}
where ${H_0} = {\omega_1}{a_1}^\dagger {a_1} + {\omega_2}{a_2}^\dagger
{a_2}$ describes the free Hamiltonian and $V = g({a_1}^\dagger {a_2} +
{a_2}^\dagger {a_1})$ is the interaction or coupling term under rotating
wave approximation (RWA) with strength of the order of $g$. $\omega_i
(=2\pi \nu_i)$ are the measure of the frequencies $\nu_i$ and $a_i
(a_i^\dagger)$ are the annihilation (creation) operators for the two
modes. Choosing the central energy of the oscillators $\hbar \omega_0
(\omega_0 ={{\omega_1 +\omega_2}\over 2})$ to be zero and defining $\delta
={{\omega_1 -\omega_2}\over 2}$, the above equation reduces to
\begin{equation}
H = \delta ({a_1}^\dagger {a_1} - {a_2}^\dagger {a_2}) + g({a_1}^\dagger {a_2} +
{a_2}^\dagger {a_1})
\end{equation}
\end{mathletters}
    Solving the Heisenberg equation of motion for the creation and
annihilation operators for the two modes and the above hamiltonian, we
find their time evolution to be
\begin{mathletters}
\begin{equation}
\left[ \begin{array}{c}
{a_1} (t) \\ {a_2} (t) 
          \end{array} \right] = 
\left[ \begin{array}{c}
{\cal A}_1 ~~~ {\cal A}_2 \\
{\cal A}_2 ~~~ {\cal A}_1^\ast 
          \end{array} \right]
\left[ \begin{array}{c}
{a_1} (0) \\ {a_2} (0) 
          \end{array} \right]
\end{equation}
and
\begin{equation}
\left[ \begin{array}{c}
{a_1^\dagger} (t) \\ {a_2^\dagger} (t) 
          \end{array} \right] = 
\left[ \begin{array}{c}
{\cal A}_1^\ast ~~~ -{\cal A}_2 \\
-{\cal A}_2 ~~~ {\cal A}_1 
          \end{array} \right]
\left[ \begin{array}{c}
{a_1^\dagger} (0) \\ {a_2^\dagger} (0) 
          \end{array} \right]
\end{equation}
\end{mathletters}
where, ${\cal A}_1 = \cos (\Omega t) - i{\delta \over \Omega} \sin 
(\Omega t)$ and, ${\cal A}_2 = - i{g\over \Omega} \sin (\Omega t)$ with 
$\Omega = \sqrt{\delta^2 + g^2}$.\\

    To calculate the different matrix elements of the observables showing
nonclassical properties for the two modes we define our product oscillator
state
\begin{equation}
\vert \Psi \rangle = \vert \psi_1 \rangle \otimes \vert \psi_2 \rangle =
\vert \alpha_1 ,\xi \rangle \otimes \vert \alpha_2 \rangle
\end{equation}
where the first oscillator is in squeezed coherent state $\vert \alpha_1,
\xi \rangle$ and the second one is in coherent state $\vert \alpha_2
\rangle$. $\alpha_i$ are  complex coherence parameters of the two modes 
and $\xi = r e^{i\phi}$ is the squeezing parameter of the first mode. 
Using the results of time evolution of the creation and annihilation 
operators and setting $\xi$ to be real and equal to be $r$, we calculate
the uncertainties in the quadratures in the two modes (first index is for
mode while the second index is for quadrature) for the state $\vert \Psi
\rangle$
\begin{mathletters}
\label{quadrature}
\begin{equation}
\Delta X_{11}^2 = {1\over 2} + \vert {\cal A}_1 \vert^2 \sinh r [ \sinh r + 
\cosh r \cos 2{\theta_{{\cal A}_1}}] 
\end{equation} 
\begin{equation}
\Delta X_{12}^2 = {1\over 2} + \vert {\cal A}_1 \vert^2 \sinh r [ \sinh r - 
\cosh r \cos 2{\theta_{{\cal A}_1}}] 
\end{equation} 
\begin{equation}
\Delta X_{21}^2 = {1\over 2} + \vert {\cal A}_2 \vert^2 \sinh r [ \sinh r + 
\cosh r \cos 2{\theta_{{\cal A}_2}}] 
\end{equation} 
\begin{equation}
\Delta X_{22}^2 = {1\over 2} + \vert {\cal A}_2 \vert^2 \sinh r [ \sinh r - 
\cosh r \cos 2{\theta_{{\cal A}_2}}] 
\end{equation}
\end{mathletters}
where, $\theta_{{\cal A}_i}$ are the arguments of ${\cal A}_i$
respectively. The terms in the square brackets can in general be negative
according to the choice of parameters. We have plotted their time
evolution in fig.2 for $\vert \alpha_1 \vert =\vert \alpha_2 \vert
=5.0$ and, $g={\delta \over {10}}$ with different $r$ to show transfer of
squeezing.  Figs. 2(a-b) shows oscillation in the uncertainties of the
quadratures with the frequency $\Omega$t. Without interaction this would
have been 2t \cite{henry}. So, by controlling the interaction strength
and/or the frequency difference between the oscillator the first
oscillator can be controlled. The plots in figs. 2(c-d) show that the
squeezing is acheived in one quadrature of the second oscillator while the
other quadrature is antisqueezed throughout the time which is not seen in
single mode squeezing. However, they oscillate in time with the same
frequency $\Omega$t and come back to the coherent state periodically. In
the case of single mode squeezing the uncertainty ellipse rotates in the
phase space in time, but here the uncertainty circle deforms to ellipse
and return back to circle in time. The amount of squeezing increases
with the degree of squeezing of the first mode ($r$) as expected. The
squeezing can be generated in the other quadrature of the second mode by
simply setting $\xi = -r$ ($\phi = \pi$).\\

     We have also calculated the mean and variances of the number
operators for both the modes. The mean numbers are given by 
\begin{mathletters}
\begin{equation}
\langle n_1 \rangle = \vert {\cal A}_1 \nu \vert^2 + \vert {\cal A}_1 
\alpha_1 + {\cal A}_2 \alpha_2 \vert^2  
\end{equation}
\begin{equation}
\langle n_2 \rangle = \vert {\cal A}_2 \nu \vert^2 + \vert {\cal A}_2 
\alpha_1 + {\cal A}_1 \alpha_2 \vert^2 
\end{equation}
\end{mathletters}
It is trivial to put the expressions of ${\cal A}_i$ and show that the
mean numbers oscillate confirming the transfer of number of photons i.e.
energy from one mode to the other. However, the total number operator
$\hat{N}$ (=$\sum a_i^\dagger a_i$) remains invariant over time. Due to the
interest about the statistics of the photon numbers in both the modes we
have calculated the variances in the number as 
\begin{mathletters}
\begin{eqnarray}
\Delta n_1^2 = [\vert {\cal A}_1 \nu \vert^2 + \vert {\cal A}_1 \alpha_1 + 
{\cal A}_2 \alpha_2 \vert^2 ] &+& \vert {\cal A}_1 \vert ^2 \sinh r \left[ 
\sinh r \right. \nonumber \\
&-& \left. 2\vert {\cal A}_1 \alpha_1 + {\cal A}_2\alpha_2 \vert^2 \{ \sinh r 
- \cosh r \cos 2\theta_{{\cal A}_1 \alpha_1 + {\cal A}_2 \alpha_2} \} \right]
\end{eqnarray}
\begin{eqnarray}
\Delta n_2^2 = [\vert {\cal A}_2 \nu \vert^2 + \vert {\cal A}_2 \alpha_1 + 
{\cal A}_1 \alpha_2 \vert^2 ] &+& \vert {\cal A}_2 \vert ^2 \sinh r \left[ 
\sinh r \right. \nonumber \\
&-& \left. 2\vert {\cal A}_2 \alpha_1 + {\cal A}_1 \alpha_2 \vert^2 \{ \sinh r 
- \cosh r \cos 2\theta_{{\cal A}_2 \alpha_1 + {\cal A}_1 \alpha_2} \} \right]
\end{eqnarray} 
\end{mathletters}
where, $\theta_{{\cal A}_1 \alpha_1 + {\cal A}_2 \alpha_2}$ and
$\theta_{{\cal A}_2 \alpha_1 + {\cal A}_1 \alpha_2}$ are the phases of
$({\cal A}_1 \alpha_1 + {\cal A}_2 \alpha_2 )$ and $({\cal A}_2 \alpha_1 +
{\cal A}_1 \alpha_2 )$ respectively. Note that the number uncertainty of
the first mode is no longer time-independent as in the case of a single
squeezed radiation mode. Though the number uncertainty can show squeezing,
but we are not interested in that at present as it has importance in
context of phases of the modes. \\

    The important features in the calculation of the number variables is
not in the result of the mean or the uncertainties, but in the difference
between them (we call it bunching parameter $\langle {{\cal B}_i}\rangle
= \Delta n_i^2 - \langle n_i \rangle$), which is a measure of the 
statistics of the number of photons in the different modes. If $\langle
{\cal B}_i \rangle$ is positive or negative, then the statistics of that
mode ($i$) follows super- or sub-Poissonian statistics and that mode is
called to be bunched or antibunched. In figs.~3(a-b) we plot the time 
evolution of the bunching parameter ($\langle {\cal B}_i \rangle$). The
bunching parameter of the first mode (fig.3a) is no longer a constant 
difference between constant variance and constant mean, but oscillates 
in time. The oscillator was initially chosen to be bunched ($\langle 
{\cal B}_1 \rangle > 0$) and the oscillation is small enough to maintain
it to be bunched. But the bunching parameter of the second mode shows 
both antibunching ($\langle {\cal B}_2 \rangle < 0$) and bunching 
($\langle {\cal B}_2 \rangle > 0$) as $\langle {\cal B}_2 \rangle$ 
oscillates over time. The amount of antibunching is seen to be more than
bunching for the second mode. Also the time spent as antibunched is 
greater than the time spent as bunched.\\

    In conclusion, we have calculated the noises in the quadratures and 
the bunching parameter for both the modes as a function of time to show
that the sharing of the non-classical properties of a squeezed harmonic
oscillator to a coherent (classical) one is possible when they are 
coupled. The procedure is supported by an experimental scheme. The 
experimental situation can also be acheived where two radiations of 
different frequencies interact through a medium consisting of three-level
atoms. The energy difference of the allowed transitions of the atom should
be resonant to that of the radiations or vice versa. The interaction 
effect generated by the atoms is totally described by the interaction 
strength $g$. In any of these cases, the system can be described by the
coupling hamiltonian of eqn.2. The transfer of squeezing clearly depends
on the interaction strength or in other words on the ratio of it with the
frequency difference between the oscillators. The squeezing generated in
the second mode is shown to be different in nature than single mode 
squeezing. We have also shown that the photon statistics of the affected
mode can be made sub- or super-Poissonian to generate antibunched or 
bunched photon beam. The procedure given here could help in generating 
and/or controling nonclassical states of radiation for applications in
quantum optics.

\end{document}